\documentclass[12pt]{iopart}
\usepackage{epsfig}

\begin{document}

\title[Double Degenerate Stars]{Double Degenerate Stars}

\author[]{
LUO Xin-Lian ({ÂÞÐÂÁ¶}\hspace{20 mm}),$^{1,2}$  BAI
Hua({°×»ª}\hspace{20 mm})$^{1}$ and ZHAO Lei({ÕÔÀÚ}\hspace{20
mm})$^{1}$}

\address{
{$^1$ Department of Astronomy, Nanjing University, Nanjing,
210093, China}\\
 $^2$ Center for Gravitational Wave Astronomy, University of Texas
at Brownsville, TX 78520, US}

\begin{abstract}
Regardless of the formation mechanism, an exotic object, Double
Degenerate Star (DDS), is introduced and investigated, which is
composed of baryonic matter and some unknown fermion dark matter.
Different from the simple White Dwarfs (WDs), there are additional
gravitational force provided by the unknown fermion component inside
DDSs, which may strongly affect the structure and the stability of
such kind of objects. Many possible and strange observational
phenomena connecting with them are concisely discussed. Similar to
the normal WD, this object can also experience thermonuclear
explosion as type Ia supernova explosion when DDS's mass exceeds the
maximum mass that can be supported by electron degeneracy pressure.
However, since the total mass of baryonic matter can be much lower
than that of WD at Chandrasekhar mass limit, the peak luminosity
should be much dimmer than what we expect before, which may throw a
slight shadow on the standard candle of SN\,Ia in the research of
cosmology.

\end{abstract}

\ead{\mailto{xlluo@nju.edu.cn }}

\vspace{.3in}


\noindent It is widely believed that White Dwarfs (WDs) are the end
stage of the low or the intermediate mass stars' evolution and are
very dense objects. Opposite to so many uncertainties under extreme
densities inside Neutron Stars (NSs), the physical bases are much
simple and well established. From observational side, there are
abundant WDs in our Galaxy due to the high frequent birth rate for
their progenitor stars and the relatively slowly cooling efficiency
after they were born. This is why the WD's structure is treated as
one of the best understood areas of astrophysics now and an
excellently educational stuff in so many astronomy textbooks, but
they are still interesting objects for scientists and catch much
attention at all times especially after Sloan Digital Sky Survey
(SDSS) data release 4 (DR4) catalog of WDs,
$^{\cite{Adelman-McCarthy}}$ which provides quite well opportunity
for a detailed comparison between theoretical models and
observations.

From theoretical viewpoint, WDs and NSs are all special cases of
Fermion stars (FSs), in which the inward self-gravity is balanced by
the degeneracy pressure of fermions. In fact, one-component FSs
except WDs and NSs are theoretically oversimplified and well studied
celestial objects for a quite long time. Their maximum mass can span
several orders of magnitude and which make them be one kind of
potential replacers of black holes. Although the observational
evidences for black holes exist have so far been quite abundant and
becoming even more strong recently, the existence of them is still
and will continue to be an hotly debated question in the following
decades before people can test GR theory through observations of
material falling into black holes as the next generation space-borne
plans will do. Instead of putting a supermassive black hole in the
center of galaxies, some peculiar objects such as FSs were
constructed theoretically and some of them can also be confronted
with many known observation constraints quite well. One famous model
among them is the extended Neutrinos Star composed of degenerate
heavy neutrinos,$^{\cite{Bilic}}$ this object has neither an event
horizon nor a singularity, with shallow potential and is benefit to
explain the soft spectrum radiation of accreting baryonic matter.
However, the 15.2-year orbit measurements of S2 surrounding a dark
object around SgrA* near the Milky Way center seem exclude the
possibility of the massive, degenerate FS in our galaxy center since
the strongly constrain form the central density
structure.$^{\cite{Sch}}$ Anyway, even one component FSs really
exist, it would be very difficult to observe them directly, since
the composition of them entirely is dark matter, which only
participates in the gravitational and sometimes in weak interaction,
does not emit or reflect light.

In this Letter, we concentrate on an idealized celestial objects
from theoretical side, namely Double Degenerate Stars (DDSs),
sometimes which look like WDs but in fact are assumed to be composed
of the normal matter with uniform chemical composition and a sort of
unknown fermions (perhaps some dark matter composition) with mass
$m_f$. Actually, the DDSs we concerned here are some kind of
fermion-fermion stars.$^{\cite{Zhang}}$ It is necessary to emphasize
that they have normal matter surfaces, on which thermal emission due
to the cooling or some internal heating process can be observed. Of
course, researchers can also identify the characteristic spectrums
due to different chemical composition in their crust and atmosphere.
If they exist and happen to be located in binary systems, we can
directly observe them and distinguish them from the normal WDs by
their peculiar behaviour.

We shall begin our discussion with the internal structure of DDSs.
To make a simple model, we have assumed that the constituent matter
is ``cold'' (fully degenerate) gas, except the quantum pressure of
its electrons or its fermions without any interactions among them,
such as the neutronization and pyconuclear reactions at sufficiently
high density, the Coulomb corrections at low density, the
Thomas-Fermi correction and so on. Now the Equation of State (EOS)
can be expressed by some very simple functions of the dimensionless
Fermi momentum $x_{i}=p_{F}/m_{i}c$,
\begin{equation}\label{1}
\rho_{i}(x_{i})=\frac{\pi }{3} \frac{g_{i}}{2} \frac{m_i^4 c^3
}{h^3} \left[3\,x_i \sqrt{x_i^2+1} \left(2 x_i^2+1\right) - 3\,
{\rm{sinh}}^{-1}\left(x_i\right)\right] \; ,
\end{equation}
\begin{equation}\label{2}
P_{i}(x_{i})=\frac{\pi }{3} \frac{g_{i}}{2} \frac{m_i^4 c^5 }{h^3}
\left[x_i \sqrt{x_i^2+1} \left(2 x_i^2-3\right) + 3\,
{\rm{sinh}}^{-1}\left(x_i\right)\right] \; ,
\end{equation}
where $\rho_{i}$, $P_{i}$, $m_{i}$ and $g_{i}$ are energy density,
degenerate pressure, rest mass and degeneracy factor of the $i-$th
component. Latin character $i$ ($i=e$, $f$) denotes the electron or
certain fermion. The only one difference between WDs and so-called
FSs we should keep in mind is that the gravitation of WDs comes
mainly form nucleons since the charge neutral condition and the
electron's mass is about 1800 times smaller than that of a proton,
whereas the gravitation of FSs is completely come from component
fermions themselves.

If we further assume that these objects are spherically symmetric,
non-rotating, non-magnetic and in hydrostatic equilibrium, then the
problem is simple enough to deal with. The gravitational field
inside these object can be expressed by the internal metric
\begin{equation}\label{3}
ds^{2}= - c^{2} B(r) dt^{2}+A(r) dr^{2}+r^{2}
(d\theta^{2}+\sin^{2}\theta d\phi^2) \;.
\end{equation}
Combining the above EOS with the continuity equation and the
hydrostatic equilibrium equation in general relativity (i.e. the
Tolman--Oppenheimer--Volkoff (TOV) equation but for multi-component
perfect fluid $^{\cite{Olson},\cite{Yuan}}$), we can obtain the
internal structure equations for these objects.
\begin{equation}\label{4}
A(r)= \left[1-\frac{2 G m(r)}{r\, c^{2}} \right]^{-1} \;,
\end{equation}
\begin{equation}\label{5}
\frac{d \ln B(r)}{dr} = \frac{2Gm(r)}{r^{2} c^{2}} \,\left[1+
\frac{4 \pi r^{3} \sum_{i}{P_i (r)}}{m(r) \, c^{2}} \right]\,A(r)
\;,
\end{equation}
\begin{equation}\label{6}
\frac{dm(r)}{dr}=4 \pi r^{2} \left[ \rho_{p}(r)+ \rho_{e}(r)+
\rho_{f}(r) \right] \; ,
\end{equation}
\begin{equation}\label{7}
\frac{dP_{f}(r)}{dr}=-\frac{c^{2}}{2} \frac{d \ln B(r)}{dr}
\left[\rho_{f}(r)+ \frac{P_{f}(r)}{c^{2}}\right] \; ,
\end{equation}
\begin{equation}\label{8}
\frac{dP_{e}(r)}{dr}=-\frac{c^{2}}{2} \frac{d \ln B(r)}{dr}
\left[\rho_{p}(r)+\rho_{e}(r)+ \frac{P_{e}(r)}{c^{2}}\right] \; ,
\end{equation}
where $A(r)$ and $B(r)$ are the metric coefficients, $m(r)$ denotes
the ``gravitational mass" inside radius $r$, which is the mass a
distant observer would measure by its gravitational effects, for
example, on orbiting movement or on gravitational lensing. Here
$\rho_{p} = m_{p} \mu_{e} \frac{8 \pi}{3} \left(\
{h}/{m_{e}c}\right)^{-3} x_{e}^{3}$ is the mass density of proton,
$m_{p}$ is the mass of proton, $\mu_{e}$ is the mean atomic mass of
electron. In additon, the particle number confined within a sphere
of radius $r$ satisfies
\begin{equation}\label{5}
\frac{dN_{i}(r)}{dr}=4 \pi r^{2} \frac{8 \pi }{3} \frac{g_{i}}{2}
\left( \frac{h}{m_i \, c } \right)^{-3} x_{i}^{3}\,\left[1-\frac{2 G
m(r)}{r\, c^{2}} \right]^{-1/2}  \; .
\end{equation}

Figure 1 shows the Mass-Radius ($M$--$R$) relations for fully
degenerate WDs, pure FSs with different fermion mass and some DDSs
with fixed fermion number. The maximum of these curves corresponds
to the Oppenheimer--Volkoff limits for degenerate
stars.$^{\cite{Oppenheimer}}$ For WDs, the limiting value is
$R_{Ch}$ $=$ $2.64 \times 10^{-2} m_{e}^{-1} m_{p}^{-1} \mu_{e}^{-1}
\left(3 h^{3} / \pi \, c \, G \right)^{1/2}$ $\approx$ $1.02 \times
10^{3} \, {\rm km} \left( \mu_{e}/2 \right)^{-1} $, $M_{Ch}$ $=$
$0.195 \, m_{p}^{-2} \mu_{e}^{-2} \left( {3 \, c^{3} \,h^{3} } /
{\pi \, G^{3}} \right)^{1/2}$ $\approx$ $1.39 \, M_{\odot} \left(
\mu_{e}/2 \right)^{-2}$, which is the Chandrasekhar mass with
general relativity corrections. However, for FSs, the limiting
values is $R_{OV}$ $=$ $0.218 \, \sqrt{2/g_{f}} m_{f}^{-2} (3 h^{3}/
\pi\, c\, G)^{1/2}$ $\approx$ $8.09 \, {\rm{km}} \sqrt{2/g_{f}}
\left({m_{f}c^{2}}/{\rm{GeV}}\right)^{-2}$, $M_{OV}$ $=$ $0.025 \,
\sqrt{2/g_{f}} m_{f}^{-2} (3 c^{3} h^{3}/ \pi\, G^{3})^{1/2}$
$\approx$ $0.627 \, M_{\odot} \sqrt{2/g_{f}}
\left({m_{f}c^{2}}/{\rm{GeV}}\right)^{-2}$, which strongly depend on
the fermion mass. The curves to the right of the maximum are stable
branch, where the radius decrease with increasing mass as we known
well in degenerate stars, while those left from the maximum
represent unstable configurations, will suffer gravitational
collapse by some unstable modes and will finally spiral into certain
points on $M$--$R$ plot as the central particle number density tends
to infinity. In Newton's theory of gravity, the upper mass limit of
WDs, i.e. Chandrasekhar mass limit, is $M_{\rm Ch}$ $\simeq$ $1.457
\cdot (2/\mu_{e})^{2} M_{\odot}$, as the center number density tends
to infinity and the radius tends to zero. Instead, the critical
radius $R_{Ch}$ for stability can be reasonably settled in the
framework of general relativity (GR).

Figure 1 also shows that the largest dimensionless surface potential
($ 2\,G M / R c^{2}$) of equilibrium FSs ($\sim 0.23$, does not
depend on the details characters of the fermion) can be much higher
than that of WDs ($\sim 4.0 \times 10^{-3}\left( \mu_{e}/2
\right)^{-1}$), which implies that general relativity is more
important in determining the structure of FSs as their central
density is high enough.

Since the mean distance between stars in typical galaxy should be
$\sim 1$ pc and the compact objects people observed in the X-ray
binary systems are merely of a few solar masses (typical value for
stellar mass black hole candidates $\sim 10\,M_{\odot}$), the DDS we
constructed in stellar level should subject to these constraints.
Furthermore, because there is no dissipation of energy due to
friction and no effectively viscous processes to transport angular
momentum for those unknown particles, the significant mass growth of
DDS itself in relatively short time duration seems impossible to
realize by accretion of dark matter. Thus, we can simply but
appropriately assume that the DDS may satisfy the condition with
conserved fermion number. Considering such number conserved pattern
and further assuming that they are composed by $0.1$\,GeV fermions
(the low limit for the mass range of Weakly-Interacting Massive
Particle, which is selected somewhat arbitrarily from almost
completely uncertain region at present), we can obtain the $M$--$R$
relations for DDSs for fixed fermion mass but with different fermion
number, as shown in Figure 1.

The calculated results from the numerical solution of the structure
equations show that DDSs with smaller electron number are composed
of a double component core and a pure fermion envelope. As normal
matter increases, the core size will increase (the dotted lines with
circles or triangles in Fig.\,1), larger gravity offered by normal
matter inside the core may act on the fermions in the outer envelope
and need more pressure to balance its structure and causes the
invisible fermion surface of DDS shrink progressively (the solid
lines with circles, triangles or asterisks in Fig.\,1). After that,
we will confront with two kinds of situations. Firstly, if there are
a sufficient number of fermions inside DDSs, the objects may always
have pure fermion envelope and their structure are always dominated
by fermion component (such as $N_{f} = 1.25 \times 10^{57}$ and
$1.12 \times 10^{58}$ curves in Fig.1). Secondly, if fermions are
not too many, as baryonic matter infuse in, the visible normal
matter surface will gradually grow up and eventually exceed the
invisible fermion surface (the turn off point $\rm {C}$ in Fig.\,1).
After that, a seemed unstable $M$-$R$ relation branch appears in
Fig.\,1, which corresponds to the transition process from fermion
dominated ($\rm {C}$) to electron dominated ($\rm {D}$). Then the
DDSs' structure are mainly determined by the normal baryonic matter
and looked more like WDs.

Simply according to the depiction of equilibrium configuration of
degenerate stars in the textbook, you might easily come to
the``common sense" conclusions that DDSs will loose their stability
from $\rm {C}$ to $\rm {D}$ since the radius increase as the mass
increase, whose behaviour seems more like those self-bound strange
stars. As we know, the self-contained stability analysis especially
with larger nonlinear perturbations acting on the equilibrium
solution is very complicated, highly model dependent without general
procedure can be devised and goes far beyond the capability of our
recent work. For simplicity, our discussion shall be restricted to
linear stability analysis.$^{\cite{Jin}}$ We find that DDSs even in
the transition state from $\rm {C}$ to $\rm {D}$ are in the
minimum-energy configuration with $dE/dn_{c,e}=0$ and
$d^{2}E/dn_{c,e}^{2} > 0$, where $E$ is the total energy of the
system, which means that DDSs can maintain stable before they arrive
at the maximum mass (as shown in Table.\,1) just form the view of
dynamic. However, there is one more complication that needs to be
kept in mind, it is about the applicability of the EOS we used. When
the central density for baryonic matter is high enough, the Fermi
energy of the electrons may excess the threshold for the inverse
$\beta$-decay reactions, then the $\beta$ equilibrium conditions
between the chemical potential of nucleons and electrons,
$\mu_{n}=\mu_{p}+\mu_{e}$, even the more practical nuclear EOS
should be considered. To some extent, the EOS we adopted here is too
simplified.

Table.\,1 lists some physical parameters for the maximum mass DDSs
under some selected fermion number $N_{f}$ and mass $m_{f}$. Because
of the enormous mass range spanned for fermion and the completely
free input value for fermion number, their reasonable analysis
requires a deep physical understanding of the nature of Dark Matter
and the structure formation for this kind of objects over a vast
range of parameter space. We just give some demonstrations and
reveal some important properties of these objects here. For fixed
fermion mass DDSs, such as $m_{f} = 0.1$\,GeV, the first to fourth
rows in Table.\,1: As the fermion number increases, you can see that
the maximum and permissible baryonic number inside these objects
will decrease, i.e. the total rest mass for normal matter will
reduce, besides the normal matter component will be buried more
deeply inside stars. The letter $\rm {f}$ and $\rm {e}$ appearing in
fourth column represent whether fermion or electron component only
exist at core. In addition, DDSs with the total rest mass of fermion
being 1$M_{\odot}$ are taken as examples from fourth to sixth rows
in Table.\,1. It is clear that the allowable maximum baryonic number
will become smaller and the normal component will be compressed more
deeply in the centre as the composed fermion becomes more heavier.
As a demonstration, the metric coefficients of two maximum mass DDSs
are plotted in Fig.\,2, which can be used to construct the Newtonian
gravitational scalar potential ($\sim {\rm ln}B(r) /2$) in weak
field limit.

Strictly speaking, there is no relationship between the evolution
and the $M$--$R$ relation for a given fermion number DDS, the
following descriptions just help us to understand. We just imagine
that DDSs originate from innate fermion star seeds formed at very
early universe and evolve just by accreting baryonic matter. At the
beginning, any gas or dust near or bumped into the innate seed tends
to be pulled into them. Friction within the accreting material
causes it to lose mechanical energy, spiral and sink into the deep
center. As matter sinks in, concentrates and compresses together
continuously, the core density increases remarkably and the core
temperature also increases simultaneously due to the release of
gravitational potential energy. If the total mass for the sinked gas
is large enough and if there is no suitable and efficient cooling
process, the central temperature may keep on increasing and
eventually reach the critical value at which hydrogen burning can
ignite. The normal matter core may arrive at some evolutionary
stage, whose behaviour will be quite similar to the main sequence of
normal stars and is supported against gravitational contraction by
the outward thermal pressure provided by the nuclear reactions. We
plan to study such a kind of objects, give more detail and
quantitative descriptions in our future work. In this study, we just
concentrate on DDSs, completely ignore the temperature contribution
and merely treat the sunk normal matter core as zero-temperature
degenerate gas. In addition, to our knowledge, the influence of
rapid rotation, strong magnetic field, finite temperature, the
coulomb corrections at low density and the neutronization and
pyconuclear reactions at high density can remarkably affect the
internal structure of the DDSs, we should gradually include them in
our continuous work.

Finally we give some observable predictions for such a kind of
objects.

(1) Since there is additional gravitational force provided by the
unknown fermion component inside DDSs, more electron degenerate
pressure are needed to maintain the structure. Therefore, one
distinguishing characteristic of DDSs is that they must have a
smaller visible radius compared with corresponding WDs. Thus, we can
provide another model instead of strange dwarfs to explain some
strange WDs' combined observations, which appear to have
significantly smaller radii than that expected for a standard
electron degenerate WD EOS.$^{\cite{Mathews}}$

(2) As we know, the leading model for type-Ia supernova (SN\,Ia) is
still degenerate thermonuclear explosion of a accreting
carbon-oxygen WD in a closed binary system as WD's mass grows to the
Chandrasekhar Mass. DDS we introduced here can also experience such
a kind of explosion when its mass exceeds the maximum mass
permissible by electron degeneracy pressure. Since its maximum mass
of baryonic matter should be smaller than Chandrasekhar mass limit
of WD, moreover the normal matter now is situated in a more deep
potential well, the corresponding binding energy of DDS is much
larger than that of WD with the same mass, the production of
radioactive nuclide $^{56}$Ni (determines the peak luminosity) and
the total kinetic energy after the explosion (determines the
expansion time scale) should be much different to that of standard
SN\,Ia, which may throw a slight shadow on the standard candle of
SN\,Ia in the research of cosmology. Thus, it is worth rechecking
and simulating the Phillips Relation (light curve width-luminosity
relationship for SN\,Ia) for such a kind of objects in a near
future.

(3) Despite the mass for baryonic matter should be smaller than
Chandrasekhar mass, considering the invisible fermion component, the
gravitational mass of DDS can be much larger than the upper mass
limit for a WD even for an NS ($\sim 3.1 M_{\odot}$). If DDSs really
exist, we hope to find them in some binary systems.

Acknowledgments: This research was supported by the National Natural
Science Foundation of China under Grants No 10221001. LUO Xin-lian
would like thank the hospitality of Center for Gravitational Wave
Astronomy at UTB, where some of the work was performed. The authors
wish to thank an anonymous referee for his valuable suggestions to
our work.

\hspace{25 mm}

\noindent Figures and Table Caption

\hspace{15 mm}

\noindent Fig1. {The Mass-Radius relations for fully degenerate
White Dwarfs, pure Fermi Stars with different fermion mass and some
Double Degenerate Stars with fixed fermion number. The Straight
lines with slope $1$ at top left is the black hole limit.}

\hspace{15 mm}

\noindent Fig2. {Metric Coefficients inside DDSs.}

\hspace{15 mm}

\noindent Table1. {The physical parameters for some maximum mass
DDSs.}

\newpage

\begin{figure}[t]
\begin{picture}(190,220)
\put(0,0){\includegraphics{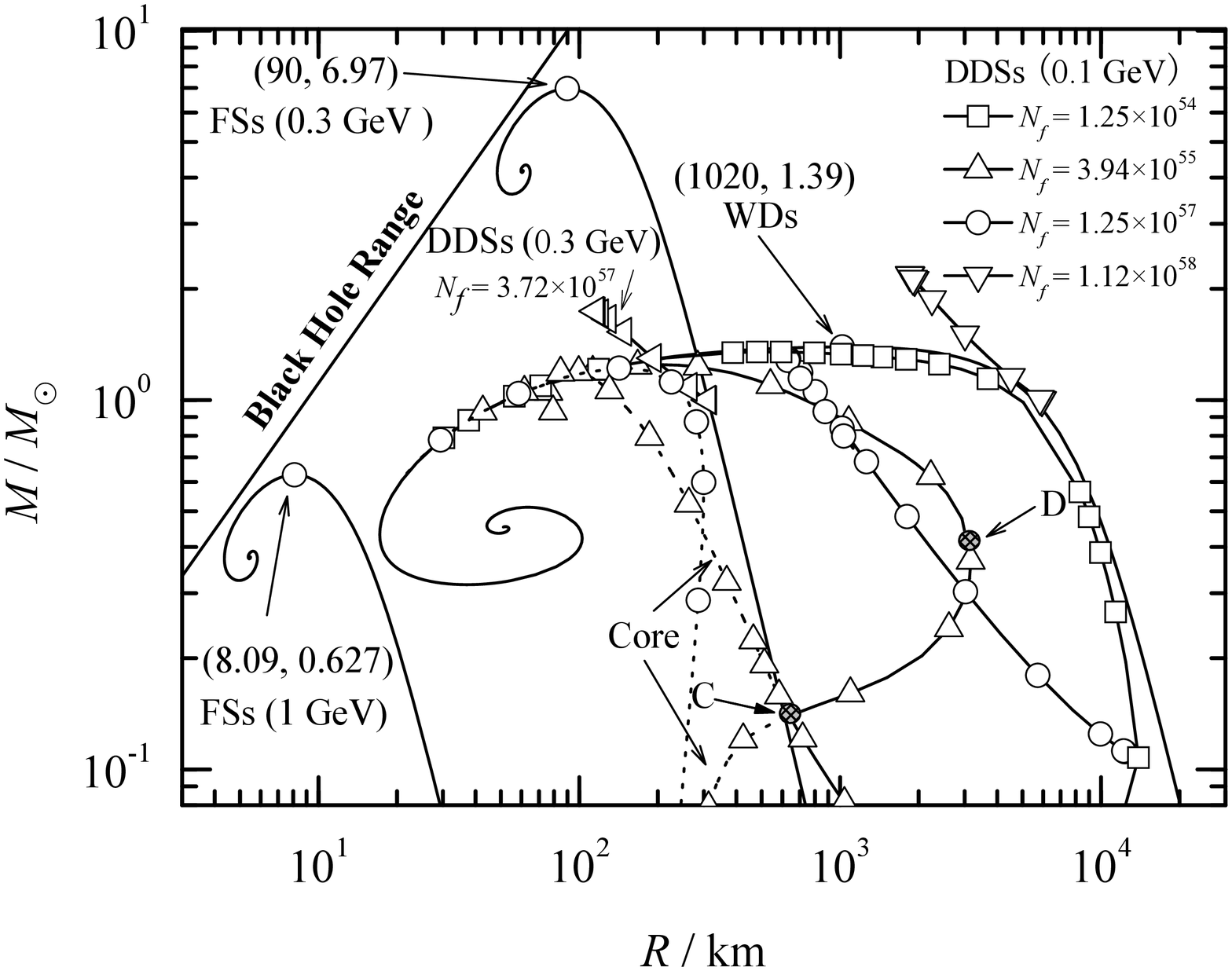}}
\end{picture}
\caption {
}
\end{figure}\label{fig1}

\newpage

\begin{figure}[t]
\begin{picture}(190,220)
\put(0,0){\includegraphics{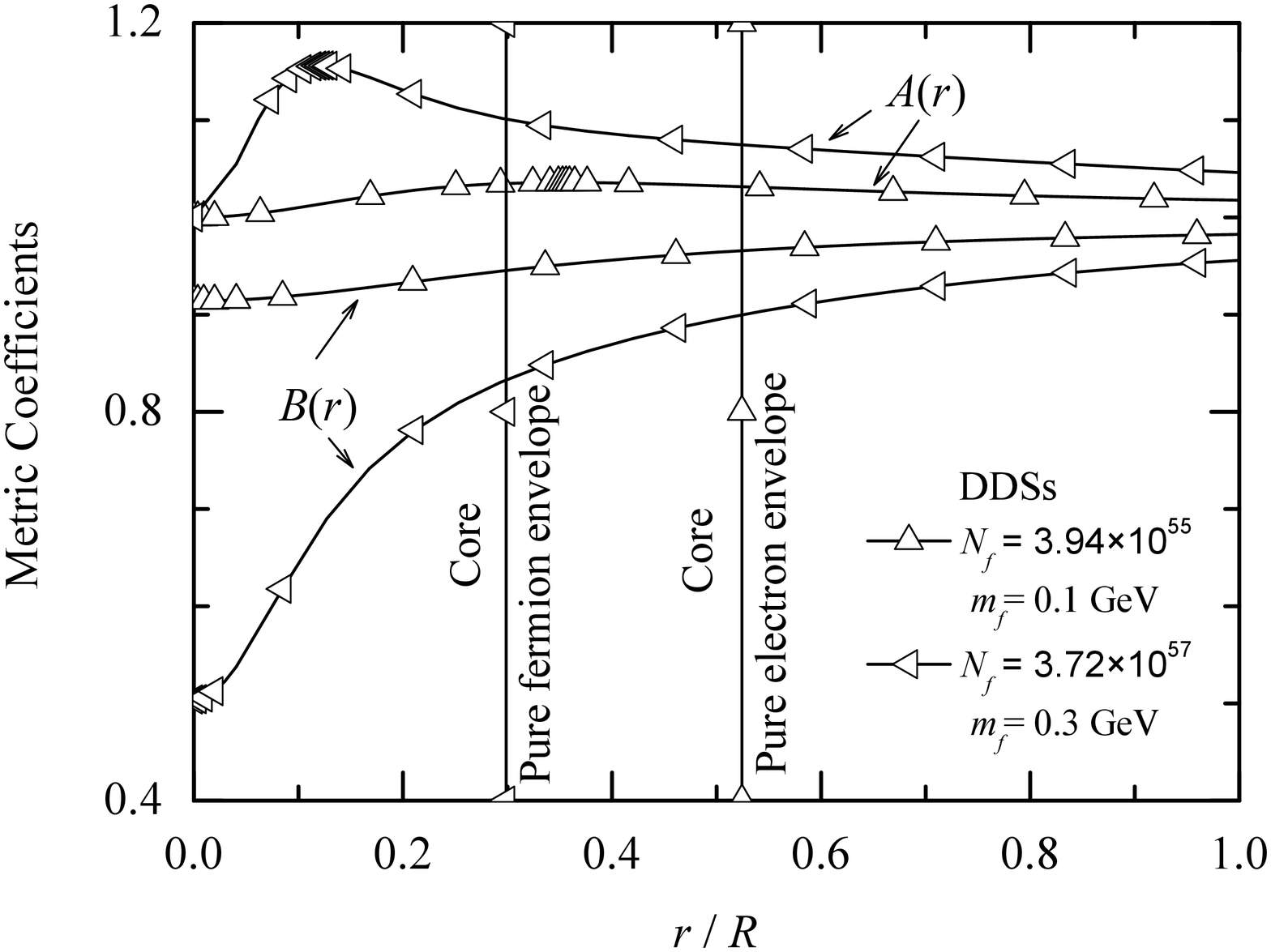}}
\end{picture}
\caption {
}
\end{figure}\label{fig2}

\begin{table}[htbp]
\caption {
}

\center\begin{tabular}{cc|cccc} \hline
$m_{f}$(GeV)               & $N_{f}$               &  $N_{e}$($/ 10^{56}$)             & Core (km)     & Radius (km)       & Mass ($M_{\odot}$) \\
\hline

$0.1$&  $1.25 \times 10^{54}$ & 8.02 &  f 90.9        & 537              & 1.35\\

$0.1$&  $3.94 \times 10^{55}$ & 7.38  &  f 111        & 209               & 1.25\\

$0.1$&  $1.25 \times 10^{57}$ & 7.06 &  e 142        & 627               & 1.30\\

$0.1$&  $1.12 \times 10^{58}$ & 7.03 &  e 135        & 1850              & 2.18\\

$0.01$&  $1.12 \times 10^{59}$ & 8.28 &  e 934        & $7.78 \times 10^{5}$ & 2.39\\

$0.3$&  $3.72 \times 10^{57}$ & 4.46 &  e 34.5        & 115 & 1.74\\

\hline
\end{tabular}
\label{tab1}
\end{table}

\end{document}